# COSRE: Community Exposure Risk Estimator for the COVID-19 Pandemic

Ziheng Sun[1]

Center for Spatial Information Science and Systems

George Mason University

**Abstract**: Due to the complexities of virus genotypes and the stochastic contacts in human society, it is a big challenge to estimate the potential risks of getting exposed to a widely spreading virus. To allow the public to be aware of the exposure risks in their daily activities, we proposed a birthday-paradox-based probability model and implement it into a web-based system named COSRE for estimating community exposure risks of public gathering during a pandemic. The exposure risk means the probability of people meeting potential COVID hosts in public places like grocery stores, gyms, libraries, restaurants, coffee shops, office suits, etc. The model has three inputs: the real-time potential active patients, the population in local communities, and the customer counts in the room. With COSRE, people can explore the possible impacts of the pandemic by doing spatiotemporal analysis, e.g., moving through time and testing a different number of people to see the changes of risks as the pandemic unfolds. The system has the potential to advance our ability to know the accurate exposure risks in various communities and allow us to make plans with improved preparedness and precise responses for the pandemic. We use spatial analysis tools and drew the county-level exposure risks of the United States from April 1 to May 15. The experiment shows that the estimation model is very promising to assist people to be more specifically precise about their safety control in daily routine and social lives and help businesses to dynamically adjust their COVID-19 policies to accelerate their recovery.

## 1. Introduction

*1.1 Motivation*

History always repeats itself regarding the pandemic [1]. Reducing the impact of the pandemic relies on early detection and mitigation strategies that slow the early spread to allow more preparatory work to be done. However, as the pandemic is already spreading right now, it puts a burden on all of us and it is not an easy way to go back to the normal. New lifestyles spread diseases further and faster than before. New and more intense factors amplify the transmission of diseases [2]. These risks apply at least equally to densely populated areas. Environmental changes such as disruptive climate, also contribute to the spreading. With the number of cases dramatically rising in United States communities exponentially, the need for risk estimation and immediate actions is never so urgent. Although self-quarantine and self-isolation are already in place in most states for several weeks, the virus is still spreading among people in grocery stores, hospitals, community activities, private parties, etc. Considering the peak might have passed, states like

[1] E-mail: zsun@gmu.edu; Mail: 4400 University Drive, MSN 6E1, Fairfax, VA 22030

Georgia start to reopen businesses to restore the economy. To assist the reopening with key assistance information, risk assessment tools are essential to guide people to determine where and when to visit.

Reopening would be a very daunting challenge for many businesses in the United States this year [3]. The duration of the pandemic is uncertain yet and the influence could last for a long time. The current guidelines for reopening are very general and not very customizable to fit in the real scenarios of different communities. The situation of each community is different regarding the population, store types, average spent time, patients, air quality, etc. It is improper to cope with them using any one-size-for-all solution. A safer plan should be able to flexibly fit by considering specific factors like the locally confirmed cases, potential asymptomatic patients, store space, layout, people density, local population composition, commodities, age distribution, capacity limit, social distancing, temperature check, violation enforcement, etc. However, further precise recommendations are rarely available for business managers to use yet. Very few quantitative models could give estimations tailored for individual businesses to make smarter decisions on balancing between life safety and economy.

*1.2 Significance*

To find a balancing point between a full lockdown in fear of the virus and a plan of reopening businesses out of socioeconomic concerns, we urgently need a mechanism to help us assist that decision. The decision-makers to be supported include not only the policymakers but also the business managers and their customers. The system should give a customized real-time risk estimation for store managers to flexibly adapt their operation to the unpredictable pandemic. The risk estimation will also provide the customers with the information regarding their risks of visiting the businesses and increase their awareness and confidence about their next move.

To achieve that goal, this project proposes a straightforward social probability model based on birthday paradox theory to estimate the risk of people meeting at least one COVID patient, either asymptomatic or symptomatic, in public places like shopping centers, grocery stores, recreational areas, restaurants, office suite, etc. A prototype web-based system named COSRE [4] has been implemented to take zip codes as inputs and output a probability of the risk. The risk is calculated based on the real-time data collected from public resources. The purpose of the tool is to give people a reasonable quantitative estimation about their risks of being exposed to the contact of COVID-19 virus in their communities, grocery stores, gyms, restaurants, workplaces, and public recreational areas.

## 2. Related Work

Pandemics are large scale outbreaks of infectious diseases that can greatly increase mortality over a big geographic area and cause significant economic, social, and political disruption [5]. The intensive global travels and long-distance contacts in human society have greatly raised the likelihood of pandemics. One example of past pandemics is the 2003 severe acute respiratory syndrome (SARS), that results in critical threats to public health [6]. The World Health Organization compel all its member states to meet specific standards for detecting, reporting on, and responding to infectious disease outbreaks. The framework contributes to a more coordinated global response during the 2009 influenza pandemic [7]. Global public health departments look for improving the preparedness through some refined standards and responding plans to flatten the curves and reduce the deaths. The management of community risks in a pandemic needs to apply more restrictive emergency management strategies than other hazards. The objectives of risk management are to strengthen our responding capacities to contain the diseases, enable and promote linkage and integration across the governments and societies. Accurate estimation of risks is the first and fundamental step in making an effective risk management plan. Common responding practices are to limit the point of entry to reduce the possibility of virus traveling. People, communities, states, and countries should keep communicating with information and advice. Precise information provided early and often will enable the community to correctly understand the health risks they face, and will make it easier to engage them in actions to protect themselves. [8]

The COVID-19 virus was first isolated from a patient with pneumonia in Wuhan, China. Genetic analysis revealed that it is closely related to SARS-Cov and genetically clusters within the genus Betacoronavirus, subgenus Sarbecovirus [9]. Although pathology studies have been done for the COVID-19 virus, as an ongoing pandemic, there are few tools available to evaluate the real-time social exposure risks. A general classification of risk assessment is grouped into several levels: little, lower, medium, high, very high, severe. Nicholas et al proposed a conceptual full-risk-spectrum comprehensive pandemic risk management system (CPRMS) to prevent, prepare for, respond to, and mitigate the multisectoral impacts of severe pandemics [10]. It contains six institutional building blocks in the global domain: governance and leadership, sustainable financing, information and knowledge systems, human capital resources, essential commodities and logistics, and operational service delivery. But the detailed framework still needs lots of work to implement in each geographical area [10].

CDC has developed an online tool called Influenza Risk Assessment Tool (IRAT) that assesses the potential pandemic risk posed by influenza A viruses [11]. The IRAT uses 10 scientific criteria to measure the potential pandemic risk associated with each of these scenarios. These 10 criteria can be grouped into three overarching categories: "properties of the virus", "attributes of the population", and "ecology & epidemiology of the virus". A composite score for each virus is calculated based on the given scenario. The score gives the means to rank and compares influenza viruses to each other in terms of their potential pandemic risks. It is an evaluation tool and should not be used as a predictive tool [12].

To take in-time, effective and appropriate actions to slow an ongoing disease outbreak, it is vital to identify risk factors such as age, gender, occupation, and health conditions, and estimate the death risk accurately in a real-time manner. For real-time infectious risk management, Jung et al proposed an approach to do a real-time estimation of the risk of death from COVID-19 infection by inferencing using the reported cases [13]. They used the exponential growth rate of the incidence to estimate the confirmed case fatality risk (cCFR) in mainland China. Tang et al studied the reproduction number of COVID-19 and tried to get more accurate R0 by using time-dependent contact and diagnosis rates to retrain their dynamics transmission model. Castro et al presented a quantitative framework for real-time ZIKV risk assessment that captures uncertainty in case reporting, importations, and vector-human transmission dynamics [14]. Their approach is divided into three sections: (1) county-level estimates of ZIKV importation and relative transmission rates, (2) country-specific ZIKV outbreak simulations, and (3) ZIKV risk analysis. They fit a probabilistic model (maximum entropy) to build a predictive model for the virus import risk. All these researches above provide important references and inspired this work.

## 3. Method
*3.1 Risk Estimator Core Algorithm*

The idea of calculating birthday paradox probability [15] is reused here. The probability distribution is similar to the Binomial distribution [16]. The basic formula is:

$$Pr_{(p,n,a)} = \begin{cases} 1 - \frac{(p-a)_!^n}{p_!^n}, & if\ p \neq 0\ and\ n \neq 0\ and\ a \neq 0 \\ 0, & if\ p = 0\ or\ n = 0\ or\ a = 0 \end{cases}$$

where $Pr$ is the probability function of meeting someone with COVID-19, $p$ is the total community population (town, village, city, county, region, country), $a$ is the total number of potential COVID-19 cases in the area, $n$ is the number of the people in the businesses like grocery stores, shopping centers, gyms, restaurants, workplaces, recreational areas, etc [17].

The idea is first to calculate the odds of **NOT** meeting any infected person and subtract that odds from 1 to get the probability of meeting at least one infected patient in that group of people.

This model is based on several assumptions. First, it assumes that all the people in the population $p$ could visit the store with equal chance. If the chance is not equal for everybody, the $n$ and $p$ should be changed according to reality. For example, the population of Fairfax County is over a million [18]. The number $p$ should exclude the people who are unlikely to show up in the stores like children, self-quarantined people, people with restricted mobility like prisoners, etc. The potential number $n$ should also remove those deceased and recovered. The business density in the area should be used to adjust the customer predicting equation for the businesses.

*3.2 Result Interpretation*

This risk estimation can help people to react reasonably instead of panicking or just being indifferent towards the virus. The formula takes three parameters as inputs and the output is the probability of the risk meeting at least one infected person in the store. The result ranges from 0 to 1. The risk can be generally interpreted as the probability of exposure to another COVID-infected person. Besides, many useful practices could be recommended by quantitatively classifying the probability into several levels. An example classification is that: the risk less than 25% is considered as relatively safe if the social distancing principle is honored and wear a mouth mask, 25-50% means the risk is relatively serious and people should wear masks and probably gloves as well, 50-75% reflect the place is very risky and masks, gloves, eye protectors are all recommended and avoid touches with the surface of any public objects (elevator buttons, stair handrails, doorknobs, etc), >75% means highly risky and the only best thing to do is stay at home and avoid any contact with anyone in public places. This is just an example and the classification should be validated with the realistic situation and evaluate its correlation between the classification and the real exposure data collected in businesses and public areas. Based on the correlation analysis results the policymakers or the people will have a better understanding of the risk numbers and take corresponding actions to avoid being exposed in their specific county.

## 4. Results

Using the proposed algorithm, this work has calculated the risks for all the counties of the lower states, and generated a series of maps since Apr 1. The COVID data comes from John Hopkins University CSSE [19], the population data for zip codes comes from the United States Census [18]. The people in the room is assumed to be 100. The estimation is made on the county level. Fig. 1 shows four of the maps on the dates of April 1, April 15, May 1, and May 13, from which we could observe the trends of social exposure risks in the pandemic.

The map of April 1 shows that New York City, Albany in Georgia, New Orleans, Denver, Salt Lake City, and Sun Valley in Idaho become the centers with high exposure risks. On the contrary, the two states, California and Washington, where the very first COVID-19 patients were found, have relatively low risks. Overall, the epidemic centers already emerge on April 1 and started spreading among the neighboring counties. At that time, most states already have their social distancing and stay-at-home orders in place. Most people start to stay at least 6 feet from other people and avoid gathering in groups. Many people tried to stay out of crowded places and most of the mass gathering is canceled following the guidelines of CDC. However, the affected people start to show symptoms and the confirmed cases are distributed all over the states. As Fig. 1(a) displays, almost every county has non-zero social risks. But risks are still low except the first several states with COVID confirmed cases. On April 1st, the risks were highest in the New York

City metropolitan, New Orleans metropolitan, Denver metropolitan, the Blaine County in Idaho, and the Summit County in Utah. A potential Outbreak could be foreseen from these areas as the curves start to rise.

The April 15 map shows that the serious areas on April 1 got even worse after two weeks. It could be seen that the red counties become more and denser. The outbreak centers on April 1st didn't change much while the spreading areas surrounding them becomes red as well. The virus spreads faster in New York, Albany, and New Orleans than the spreading in Denver, Salt Lake City, and Sunny Valley. New York has taken the hardest hit with a big portion of the neighboring counties to New York city becoming dark red. Louisiana and Mississippi are in deep red. Most counties in the two states have a significant amount of confirmed cases comparing to their population. The three counties in northern Arizona and the northwestern counties of New Mexico start to have high risks. For example, the Navajo county in Arizona only has a population of 110,924 (2019) but found 390 confirmed cases and the population-patient ratio is very high. Besides, unlike the metropolitans, the Navajo county has fewer stores and public places, which would make it even risker if businesses reopen and people gather. Besides, the Detroit and Chicago areas start to join. However, the risk in Washington and California remains at a relatively low level comparing to the other outbreak centers in the eastern and southern U.S. Florida also shows the same pattern and only have relatively high risk in the Miami region.

The calculation of the maps in May is a little different from those in April. As suggested by physicians that most confirmed cases will either recover or die within a 30-day window, the patient number used in the May calculation will extract the number of the same day in April (one month ago). The formula is:

$$n_{may\_i} = n_{may\_i\_confirmed} - n_{april\_i\_confirmed}$$

where $i$ represents the day in the month. As May has 31 days and April only has 30 days, we will use April 30 data in the calculation of May 31. The map of May 1 shows that New York, Pennsylvania, New Jersey, and Massachusetts have high exposure risk all over their coastal counties. New York City and its associated counties are the most dangerous areas. People in those areas should wear mouth masks and keep up with social distancing principles in public places. Stay home for 14 days after traveling and self-monitor for symptoms by checking temperature twice a day and watching for fever, cough, shortness of breath, and other symptoms of COVID-19. People in those risky areas should be more cautious by avoiding any contact with vulnerable groups, like people 65 years or older, and people who live in a nursing home or long-term care facility. The other east coast states, including Maryland, Virginia, Washington D.C., North Carolina, South Carolina all start to turn darker. The inland region, such as western Pennsylvania, West Virginia, western Virginia, west of North Carolina, east fo Tennessee, has relatively lower exposure risks. On May 1, the populous counties in the middle states like Iowa, Nebraska, Ohio, Indiana, Kentucky, Minnesota, have turned dark red. But they are not spreading like New York or other metropolitans. The risks are restricted in certain counties and show no obvious signs of a wide outbreak within the entire states. Those red counties are very scattered and isolated. The reason is probably that the other less populous counties intentionally reduced their contacts with the populous counties. The connections among different counties are not as intense as those in the east coast counties. The risk in Washington, California, Nevada, Texas, and Florida, is slowly rising and remains lower than the other outbreak states.

According to the map of May 13, the riskiest areas are still on the eastern coast, however, with the sign of decreasing in some suburban counties in New York, Pennsylvania, and Georgia. The surrounding counties of Columbus in Georgia saw the risks are reducing regarding the peak might be over and the curve of New York is on the downside. The populous counties in the Mississippi downstream basin and the corn belt states have become darker. The counties in the west Oklahoma and Kansas have spreading several more, and some of them are darker red. On the other hand, the risk in the New Orleans metropolitan is much lower than two weeks ago. The risk on the western coast is relatively low. The most serious counties in northern Arizona are Navajo and Apache, and in northwestern New Mexico are San Juan county and Mckinley county. The Chicago metropolitan turns to red and is very likely to become dark red on June 1. The western coast and Florida stay nearly unchanged since the beginning of this month. The two states that need to watch out in the next two weeks are those corn belt states, including Illinois, Michigan, Ohio, Kentucky, Tennessee, Wisconsin, Iowa, Indiana, and Missouri. These states haven't clearly shown a drop in the curves and some of them have begun to reopen, which probably might contribute to a big spike in the confirmed cases.

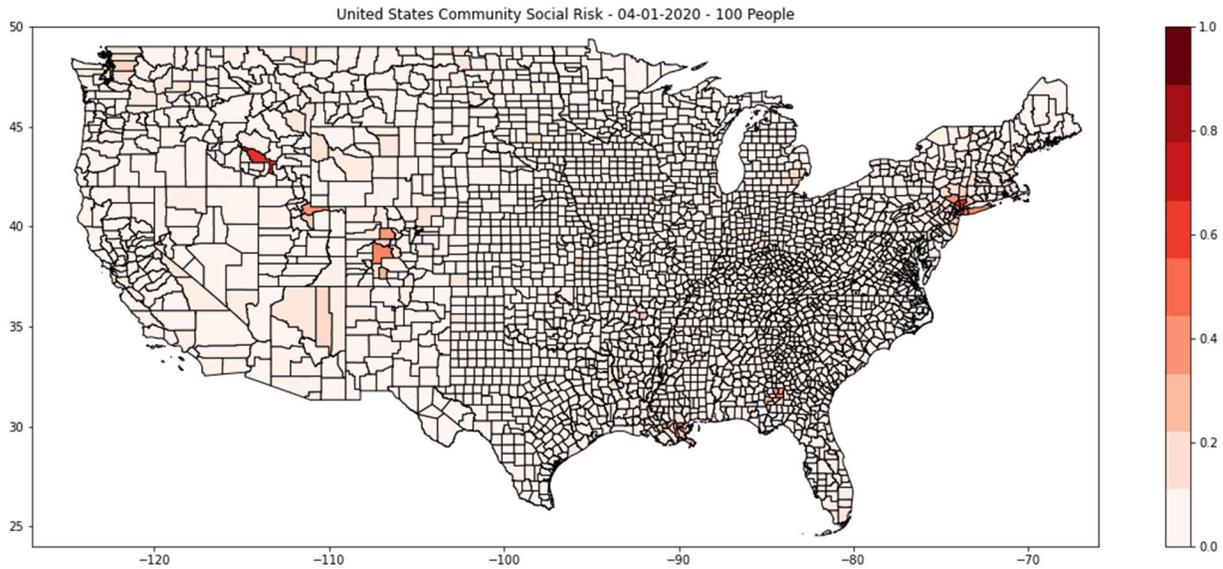

(a) April 1

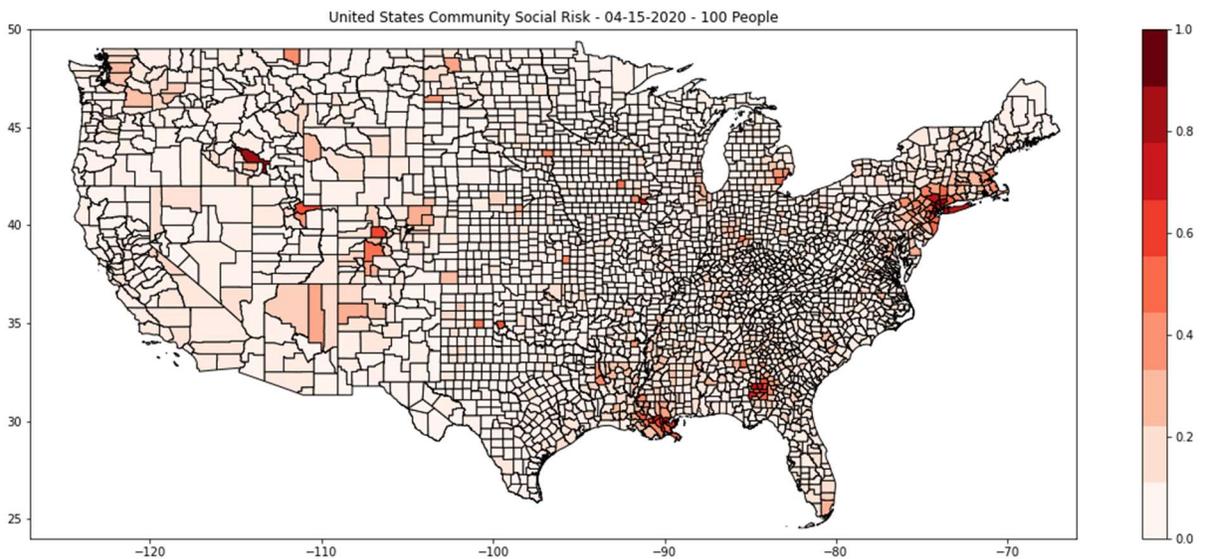

(b) April 15

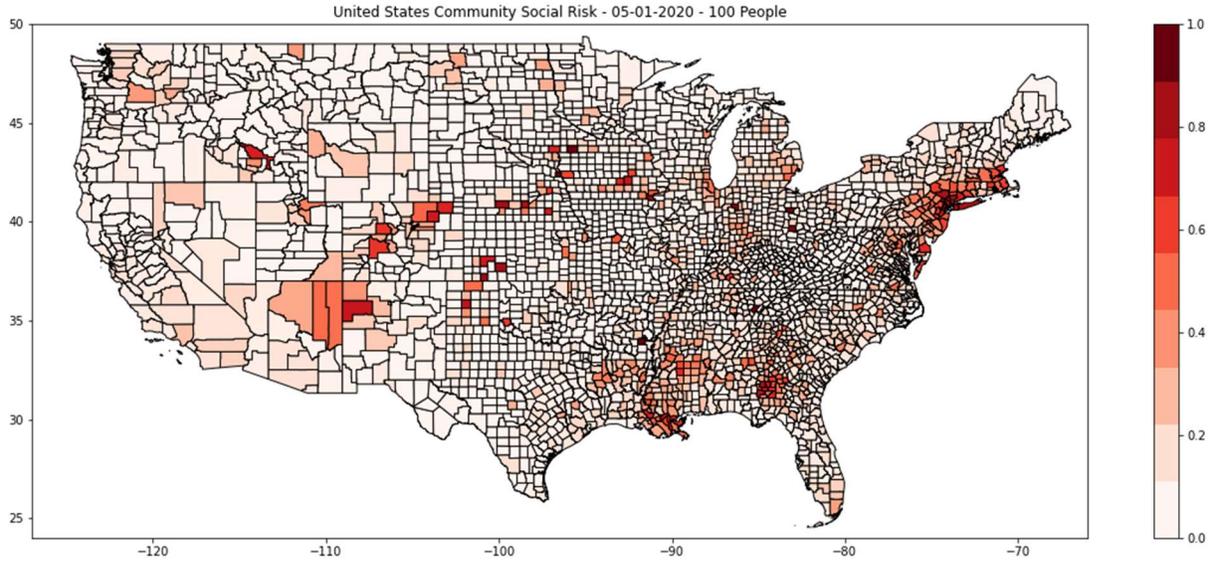

(c) May 1

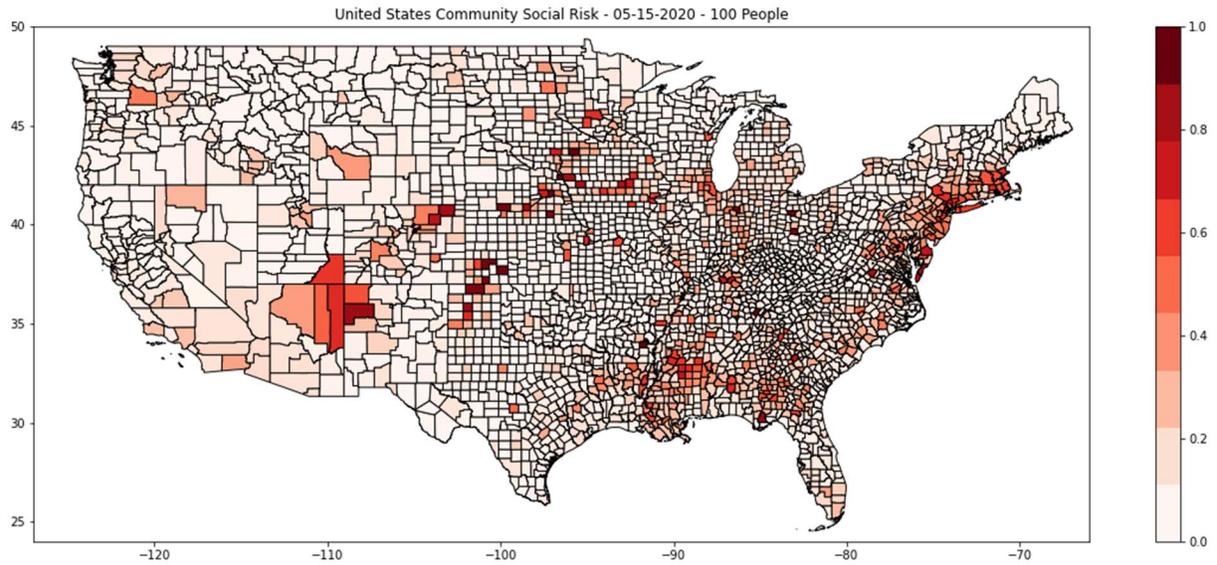

(d) May 15

Figure 1. The Community Exposure Risk of the United States on April 1, April 15, May 1, and May 15 2020

## 5. Discussion

### 5.1 Community Exposure Risk and Transmission Risk

The community exposure risk is an important contributing factor to the transmission risk. The relationship could be simply delineated using the following equation:

$$R_{transmission} = \frac{R_{exposure} \times R_{contract}}{I_{immunity}}$$

where $R_{exposure}$ is the exposure risk in community daily activities, $R_{contract}$ is the infection risk. The two risks come from two steps. The former risk estimates the possibility of people contacting the virus hosts and sources. For example, people staying at home has lower exposure risk than people who go to grocery stores. The latter risk estimates the possibility of being infected by the virus after contacting the sources. If people wear mouth masks, safety glasses, disposable gloves, and footwear, their contracting risk will be much lower than people who wear no protections. $I_{immunity}$ represents the immune ability of the person or the community to the virus. People with antibodies could be immune to certain types of viruses and their immunity is higher. The transmission risk is inversely proportional to the immune ability of the person or the entire community.

*5.2 Exposure Risk Evaluation*

It is difficult to evaluate the accuracy of the estimation. This work gives a customized quantitative estimation on the social exposure risks of people showing up in public places. The results are tailored for every community by considering their local population, potential patients, store number, potential guest flow. The purpose is to help people increase their awareness of their risks and help U.S. businesses adjust their store policy dynamically. It is also supposed to accelerate the reopening of the business while maintaining a low risk of virus contact to protect the wellbeing of both employees and customers. Upon the success of this project, people will have a clear understanding of the risk of going to the store and both customers and businesses can make wiser decisions on the fly.

However, all of these will not happen without a solid evaluation of the effectiveness and precision of the results. Model evaluation should use real data. The exposure data is rarely available right now. Only big companies like Apple and Google and some Chinese companies have used their platform and service to trace the potential patients by using the Bluetooth of their smartphones. Based on the patient tracing data, we could compare the real probability of the store with COVID patients present with the estimated risks of this work. Since the pandemic is still ongoing, such a dataset is relatively sensitive and hard to retrieve at present. Model evaluation with real-world data is our next step of work.

*5.3 Model Reliability in Reality*

A reliable model that is designed to be used in the real-world must consider many realistic factors. More refinement work should be done to improve the accuracy of the three input numbers. For example, the model assumes that the confirmed patients covers all the existing patients and are all free to move. However, in reality, there is not true. As an improvement, R0 should be used to calculate new potential COVID patient numbers and make them as close to the real numbers as possible. For the population, the model assumes that everyone in the population has the same chance of showing up in one store, which is not true in reality neither. Everyone has their preferences in grocery stores, shopping centers, coffee shops, libraries and the chance is various. One solution is to restrict the population number to the pool of potential place goers and the way to do that is getting store visiting data from SafeGraph, analyzing the age, gender, and other characteristics of the store customers, and derive a reasonable customer pool based on the regional population census. Remove people without the same chance of showing up from the population used for calculation. The third number also needs to be further improved by adjusting the real-time store visitors based on popular hours, business density, county income level, and real-time customer count prediction, etc.

## 6. Conclusion

To allow the public to be aware of the virus exposure risks in their social activities, we propose a birthday-paradox-based probability model and developed a web-based system for calculating community exposure risks for public gathering, either essential or non-essential, during a pandemic. The risks are generated based on the real-time potential COVID-19 patients, the population in local communities, and the people number in the stores. With the system, people can explore the effects of the pandemic on the population through a geographical spatiotemporal view, moving through time and testing a different number of people to see the changes of risks as the pandemic unfolds. As the virus testers become more available and cover more population, the ability to monitor and predict community-level risks amid a pandemic could be greatly augmented. The system integrates the risk estimation model, computational tools, and the analysis of evolutionary pathways, together with refinements to virus surveillance.

The experiment results show that the proposed model is promising in improving our ability to assess the risks posed to humans by the COVID-19 virus and other influenza viruses, and equipping us with improved preparedness and response for this ongoing and all the future pandemics. The application scenarios of the system are set to be in the middle of a pandemic when the tracking of patients is in place. However, due to the inaccuracy of the input three numbers and the complexity caused by the numerous realistic factors, the estimated risks might be far from reality. We are still working to improve it to be accurate and reliable enough to be practically used in the ongoing pandemic.

## Acknowledgment
The authors thank all the involved colleagues, the MIT COVID-19 Datathon organizers, the members and mentors of Team-E-006 for their valuable advice. Thanks for the COVID-19 dataset provided by John Hopkins University, the population data from U.S. Census, and the foot traffic data kindly provided by SafeGraph Inc. Thanks to all the developers of the Python libraries and tools used in this work.